\documentclass[aps,pra,reprint,superscriptaddress]{revtex4-1}
\usepackage{bm,color,ulem}
\usepackage{amsmath}
\usepackage{amssymb}
\usepackage{graphicx}
\usepackage{hyperref}
\usepackage{lineno}

\begin{abstract}
A focusing system such as a single lens or a spherical mirror imparts intrinsic transverse orbital angular momentum (OAM) to spatiotemporal (ST) coupled fields the ST intensity distribution of which presents ST covariance. This fact may greatly simplify the experimental setups used to date to impart transverse OAM. We evaluate analytically the imparted transverse OAM as a function of the focal length and the covariance. The focused fields with transverse OAM include elliptical ST vortices and rotating pulses without any ST phase singularity such as the ``lighthouse" pulse. We provide closed-form, analytical expressions for these fields valid at any propagation distance from the focusing system, which are of interest in applications such the interaction of these fields with matter. In general, focusing of ST coupled fields with intensity covariance generates mixed fields with ST vortices and rotating pulse-fronts, where one or another feature dominates depending on the input field.
 \end{abstract}

\begin{document}

\title{Transverse orbital angular momentum imparted upon focusing spatio-temporally coupled ultrashort pulses}

\author{Miguel A. Porras}
\email{miguelangel.porras@upm.es}
\affiliation{Grupo de Sistemas Complejos, ETSIME, Universidad Politécnica de Madrid, Rios Rosas 21, 28003 Madrid, Spain}
\author{Spencer W. Jolly}
\email{spencer.jolly@ulb.be}
\affiliation{Service OPERA-Photonique, Université libre de Bruxelles (ULB), Brussels, Belgium}
\date{\today}
\maketitle

\section{Introduction}

In the dynamic field of structured light, backed by major experimental achievements, orbital angular momentum (OAM) and vortex structures play a prominent role. After three decades devoted to spatial vortices and longitudinal OAM, research now expands to spatiotemporal (ST) optical vortices (STOVs)~\cite{WAN}, featuring a line phase singularity transverse to the propagation direction, and to the associated transverse OAM. These STOVs are generated using standard, two-dimensional diffractive pulse shapers with either phase plates \cite{HANCOCK_OPTICA} or spatial light modulators \cite{CHONG,CAO} placed at the Fourier plane of $4f$ systems between two diffraction gratings. Recently the use of metasurfaces has been proposed \cite{HUANG}, and methods to impart or remove transverse OAM to already formed STOVs have been demonstrated \cite{TORQUING}.

Here we show that a single focusing system such as a lens or a spherical mirror can impart transverse OAM to an OAM-free wavepacket.
The condition for the illuminating pulse to acquire transverse OAM is that its ST structure is coupled such that the intensity presents covariance between time and a transversal coordinate. The amount of transverse OAM per unit energy is evaluated as a function of the focal length and the covariance of the intensity.

With an input tilted pulse, the imparted transverse OAM produces a focusing pulse with a rotating pulse front without involving the formation of ST singularities. This is the ``lighthouse" pulse previously known by its rotating wavefronts in time at the focal plane \cite{VICENTI_PRL,QUERE_JOP,AUGUSTE_PRA}. Partially coherent rotating pulses have been described, and its transverse OAM noticed in \cite{HYDE}. Here we provide a closed-form analytical expression of rotating pulses valid at any propagation distance for coherent illumination, which is the most frequent situation in light-matter interaction experiments \cite{VICENTI_PRL,QUERE_JOP,AUGUSTE_PRA,FANG}, and evaluate its transverse OAM as a function of measurable parameters as the tilt parameter, transversal size, and focal length.

In a second example, focusing a pulse with $n$ tilted line $\pi$-steps in the phase between $n+1$ tilted intensity lobes, and carrying no transverse OAM, produces a STOV of topological charge $n$ at the focal plane. This illumination resembles the output from a $4f$ pulse shaper described in the experiments in~\cite{HANCOCK_OPTICA}, which also focuses to a STOV. Here we point out that the $4f$ pulse shaper system is not necessary since the lens already endows the light with transverse OAM as long as the covariance is present, and the multi-lobe tilted structure can be generated by other means such as spatial light modulator and a simple prism. The fact that the tilted multi-lobe structure of Hermite-Gauss shape focuses to a STOV has also been reported in \cite{CHEN,CHEN2}, where the focused field is numerically evaluated at the focal plane for $n=1$ and $2$. Here we obtain analytical expressions for the propagating pulse at any distance from the focusing system, and for arbitrary $n$, and point out that multi-lobe tilted structures with shapes other than Hermite-Gauss also focus into STOVs. In addition, we evaluate analytically the amount of transverse OAM of the focused pulse.

\section{Transverse OAM imparted by a focusing system}

We consider an optical pulse $E=\phi(x,y,t)e^{-i\omega_0 t}$ in free space that will be focused along the $z$ direction under paraxial conditions and that comprises many optical oscillations owing to its narrow-band temporal frequency spectrum about a carrier frequency $\omega_0$. We do not consider in this paper few-cycle, broadband pulses. For simplicity, $\phi$ is assumed to present ST couplings only in $x$ and $t$, so that $\phi=\psi(x,t)Y(y)$. With this choice triple integrals for the physical magnitudes of interest factorize in double integrals in $x$ and $t$ multiplied by $\int |Y(y)|^2 dy$, which will not be written, and cancel out when considering the quotient of two triple integrals. With an adequate choice of the origin of time, one can get the temporal center of pulse packet to vanish, i.e., $t_c=\int |\psi|^2 t dx dt/\int |\psi|^2 dx dt =0$ (all the integrals extend from $-\infty$ to $+\infty$). The origin of $x$ can be chosen such that the transversal center of the pulse is zero, i.e., $x_c=\int |\psi|^2 x dx dt/\int |\psi|^2 dx dt =0$.

Under the above paraxial and quasimonochromatic (many-cycle) conditions, the intrinsic OAM along the transverse $y$ direction and the energy carried by the wavepacket can be evaluated from \cite{PORRAS_PIERS}
\begin{equation}\label{TOAM}
J_y^{(i)}= -\frac{\varepsilon_0 c}{2k_0}\mbox{Im} \int \psi^\star\partial_x \psi (t-t_c) dx dt
\end{equation}
with $t_c=0$, and
\begin{equation}\label{ENERGY}
W= \frac{\varepsilon_0 c}{2}\int |\psi|^2 dx dt ,
\end{equation}
where $\varepsilon_0$ is the free space electric permittivity, $c$ the speed of light in free space, and $k_0=\omega_0/c$. The intrinsic transverse OAM is the OAM about a moving axis parallel to the $y$ direction passing permanently through the pulse center, which is conserved on propagation \cite{PORRAS_PIERS}. Here, the center is that of the intensity distribution. In \cite{HANCOCK_PRL}, the pulse center is also that the intensity distribution, and the results for the intrinsic transverse OAM coincide with those in \cite{PORRAS_PIERS} (of course other choices for the pulse center are possible \cite{BLIOKH_PRA}, as the center of the photon wave function, which leads to different values of the intrinsic OAM). It is assumed that the pulse to be focused does not carry any intrinsic transverse OAM so that the integral (\ref{TOAM}) with $t_c=0$ is vanishes.

The focusing element may focus in $x$ and $y$, or only in $x$, e.g., a cylindrical lens or mirror. Focusing in $y$ adds nothing to the discussion, but simply makes $Y(y)$ to focus remaining decoupled. We then choose a cylindrical focusing element focusing only in $x$. Its center $x=0$ is aligned with input pulse center, $x_c=0$. Otherwise the focusing system would deviate the input pulse imparting an extrinsic OAM in which we are not interested. In ideal focusing, the primary effect of the focusing element is to impart a converging spherical wave front represented by the factor $e^{-ik_0x^2/2f}$, where $f>0$ is the focal length. The second effect, for ultrashort pulses, is to introduce a pulsefront curvature described by replacing $t$ with $t-x^2/2cf$ in $\psi$.  Thus, the field immediately after the focusing system is
\begin{equation}\label{FOCUSED}
\psi_f=\psi(x,t-x^2/2cf) e^{-ik_0x^2/2f},
\end{equation}
or $\psi_f=\psi(x,t_f) e^{-ik_0x^2/2f}$, where $t_f=t-x^2/2cf$.

To the purpose of evaluating the new intrinsic transverse OAM, and the OAM per unit energy (``per photon"), we first note from (\ref{FOCUSED}) that $W_f=W$, that $x_{c,f}=0$, and that the temporal center is slightly shifted to $t_{c,f}=(\Delta x)^2/2cf$, where $(\Delta x)^2 = \int |\psi|^2 x^2 dx dt/\int |\psi|^2 dxdt$, as an effect of the pulse front curvature. Using (\ref{TOAM}) with the focused field (\ref{FOCUSED}) and with the new center $t_{c,f}$, performing the derivatives with the chain rule, changing the integration variable $t$ to $t_f$, and taking into account that $x_{c,f}=0$, one arrives at
\begin{equation}\label{CALCULUS}
\begin{split}
&J_{y,f}^{(i)}=-\frac{\varepsilon_0 c}{2k_0}\left\{ -\frac{\omega_0}{cf} \int|\psi|^2 x\left(t+\frac{x^2}{2cf}\right) dx dt
\right.\\
&+\left.\mbox{Im}\!\int \!\psi^\star\!\left[ \partial_x\psi - \partial_t\psi \left(\frac{x}{cf}\right)\right]\!\left(t\!+\! \frac{x^2}{2cf}\!-\!t_{c,f}\!\right) dxdt \right\},
\end{split}
\end{equation}
where $\psi =\psi(x,t)$ and the subindex $f$ in $t_f$ is omitted at the end. No particular assumption for the input field has been made up to this point, except that the input pulse is aligned with the focusing system.

Considering focusing, the most common situation is collimated illumination. Writing $\psi =Ae^{i\Phi}$, where $A$ and $\Phi$ are the real amplitude and phase, $\mbox{Im}\{\psi^\star \partial_x \psi\}=A^2\partial_x \Phi=0$ since the phase does not depend on $x$. Then, the input field does not indeed carry transverse OAM, and the three integral terms in the second row of (\ref{CALCULUS}) containing $\partial_x \psi$ vanish. For simplicity, and to focus on the phenomenon of interest, we will assume that the input field does not contain any temporal chirp. Similarly, $\mbox{Im}\{\psi^\star \partial_t \psi\}=A^2\partial_t \Phi=0$, and the three integral terms with $\partial_t\psi$ also vanish. Since the phase is constant, the input field only can contain ST couplings in the amplitude. With these assumptions, only the first row in (\ref{CALCULUS}) remains, and when expressed per unit energy, we obtain the intrinsic transverse OAM imparted by the focusing system as
\begin{equation}\label{FTOAM}
\frac{J_{y,f}^{(i)}}{W_f}=\frac{1}{f}\left[\frac{\int |\psi|^2 xt dx dt}{\int |\psi|^2 dx dt} +\frac{1}{2cf}\frac{\int |\psi|^2 x^3 dx dt}{\int |\psi|^2 dx dt}\right].
\end{equation}
The first term is the contribution from wavefront curvature and the second one from pulse front curvature.
The relevance of each term can be analyzed by introducing dimensionless variables $\xi=x/X_0$ and $\tau=t/t_0$, where $X_0$ and $t_0$ are characteristic half beam size and half pulse duration (as in the examples below). Then (\ref{FTOAM}) can be expressed as $(X_0t_0/f) (I_1 +\alpha I_2)$, where $I_1$ and $I_2$ are the two same quotients of integrals as in (\ref{FTOAM}) but with variables $\xi$ and $\tau$, and $\alpha = (Z_R/f)/(\omega_0 t_0)$, with $Z_R= k_0X_0^2/2$ the Rayleigh distance of the incident wave packet. The condition $Z_R/f\gg 1$ characterizes focusing without appreciable focal shift \cite{SELF,CARTER} or large Fresnel number \cite{COLIN}, and $\omega_0 t_0\gg 1$ characterizes pulses with many oscillations or long duration. Henceforth, we will choose long enough duration for the desired focusing geometry ($Z_R$ and $f$) such that $\alpha\ll 1$, whereby
\begin{equation}\label{FTOAM2}
\frac{J_{y,f}^{(i)}}{W_f}\simeq \frac{1}{f}\frac{\int |\psi|^2 xt dx dt}{\int |\psi|^2 dx dt} .
\end{equation}
In the example of Fig. \ref{Fig2} with a focusing geometry with negligible focal shift ($Z_R/f=20.8$), $\alpha=0.167$ for the duration $t_0=50$ fs. In the example of Fig. \ref{Fig3} with relevant focal shift ($Z_R/f=1.33$), $\alpha=0.011$ for the same duration $t_0=50$ fs (of course both examples are within the paraxial regime: respective divergence angles 0.4 and 0.057 deg). In addition, the second term in (\ref{FTOAM}) is exactly zero for many fields with ST couplings that preserve some symmetries, as in the two examples below, in which case (\ref{FTOAM2}) is exact.

\begin{figure}[t]
\begin{center}
\includegraphics*[height=2cm]{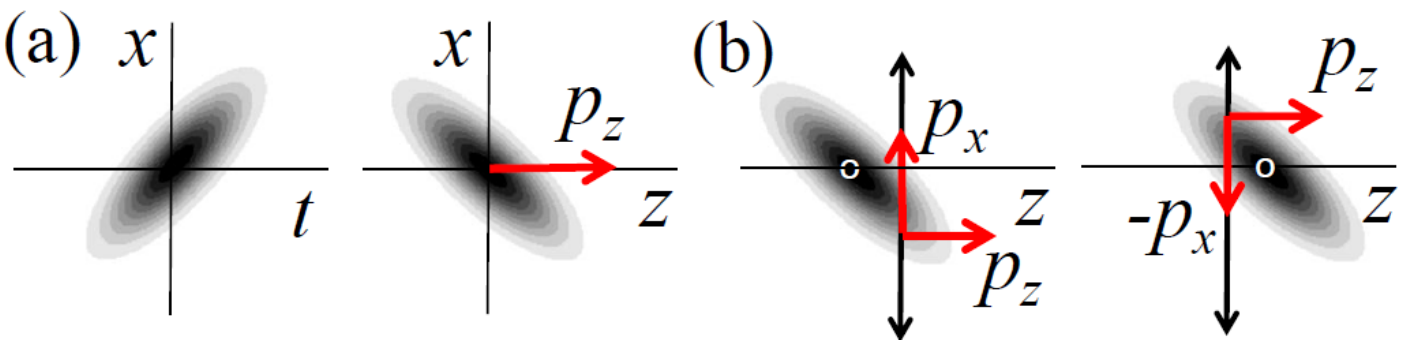}
\end{center}
\caption{\label{Fig1} (a) The pulse with positive covariance in $(t,x)$ has negative covariance in $(z,x)$. (b) The lower and upper parts of the pulse are focused at different times, imparting opposite momenta $p_x$ and $-p_x$, but the corresponding angular momenta with respect to the instantaneous pulse center (small circles) have the same sign. }
\end{figure}

Equation (\ref{FTOAM2}) is simple and looks particularly appealing conceptually. The quotient of integrals is the covariance of the intensity $|\psi|^2$ in the variables $x$ and $t$. Focusing then imparts an intrinsic transverse OAM proportional to the power $1/f$ of the focusing system if $\psi$ is a ST coupled field whose intensity distribution presents covariance. If the intensity covariates in the first and third (second and fourth) quadrants, the transverse OAM is positive (negative). A sketch of how the focusing system transmits the OAM is shown in Fig. \ref{Fig1}. Let us illustrate this result with a couple of examples of interest in experiments.

\begin{figure*}[t]
\includegraphics*[height=3.7cm]{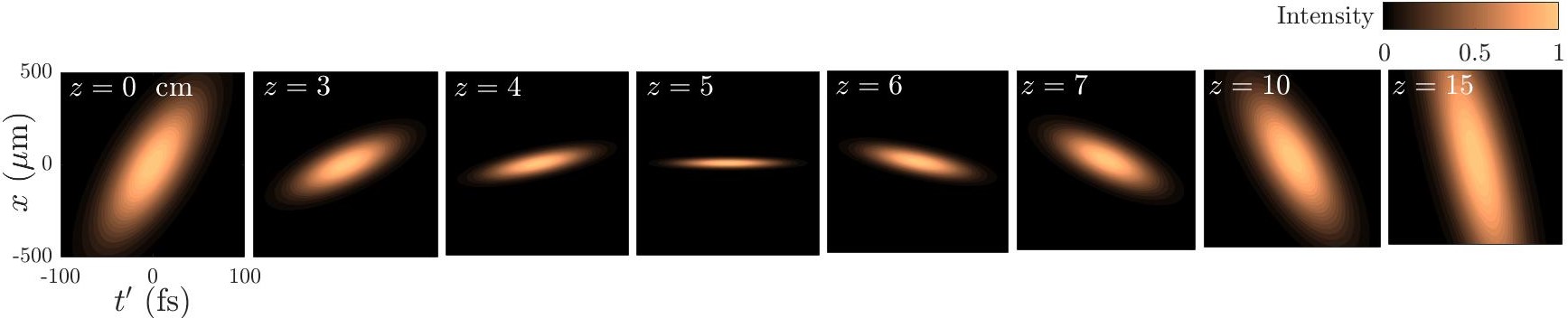}
\caption{\label{Fig2} Rotating pulse of carrier frequency $\omega_0=2.5$ rad/fs obtained by focusing the Gaussian tilted pulse $\psi=e^{-x^2/X_0^2}e^{-(t-px)^2/t_0^2}$ with $X_0=500$ $\mu$m, $t_0=50$ fs, tilt parameter $p=0.1$ fs/$\mu$m, with a focal length $f=5$ cm. The intensity is normalized to the peak intensity at each distance.}
\end{figure*}

\section{Examples}\label{Examples}

\subsection{Rotating pulse}

Let the illumination be the tilted pulse \cite{BOR93,PRETZLER00}
\begin{equation}\label{TP}
\psi = \psi_x(x)\psi_t(t-px),
\end{equation}
where $p$ is the tilt parameter, and the functions $\psi_x(x)$ and $\psi_t(t)$ are taken real. Then the illumination (\ref{TP}) does not carry transverse OAM. It is readily seen that $x_c$ and $t_c$ of the tilted pulse (\ref{TP}) are zero if the mean values of $\psi_x(x)$ and $\psi_t(t)$ are zero as functions of their respective variables. After focusing, a simple calculus from (\ref{FTOAM2}) in which many integrals factorize and cancel out, results in
\begin{equation}\label{TOAMR}
\frac{J_{y,f}^{(i)}}{W} \simeq\frac{p}{f}\frac{\int |\psi_x|^2 x^2 dx}{\int |\psi_x|^2 dx} = \frac{p}{f} (\Delta x)^2,
\end{equation}
meaning that the tilted pulse acquires an intrinsic transverse OAM proportional to the tilt parameter and to the transversal width.

With the Gaussian tilted pulse $\psi=e^{-x^2/X_0^2}e^{-(t-px)^2/t_0^2}$, the intrinsic transverse OAM per unit energy after focusing is $J_{y,f}^{(i)}/W=pX_0^2/4f$. The second term in (\ref{FTOAM}) due to pulse front curvature vanishes in this case. The manifestation of this transverse OAM is an intensity pattern that rotates about a transverse $y$ axis passing permanently through its center during propagation, as seen in Fig. \ref{Fig2}. Here, the transverse OAM does not involve the formation of any ST phase singularity, hence the phase pattern is not shown.

Since pulse-front curvature does not contribute to the transverse OAM, we can evaluate the focused field neglecting it to obtain a field with the same OAM content. We take Fresnel diffraction integral
\begin{equation}\label{FRESNEL}
\begin{split}
&\psi_f(x,t',z) =\sqrt{\frac{k_0}{2\pi i z}}\\
&\times\int dx' \psi(x',t') e^{-ik_0 x^{\prime 2}/2f} e^{ik_0(x-x')^2/2z},
\end{split}
\end{equation}
($t'=t-z/c$ is the local time) as the solution of the paraxial wave equation $\partial_z \psi = (i/2k_0)\partial_{xx}\psi$ for paraxial, quasimonochromatic (many cycle, $t_0\gg c/\omega_0$) pulses in absence of material dispersion, for focusing the tilted pulse $ \psi(x,t)=e^{-x^2/X_0^2}e^{-(t-px)^2/t_0^2}$ at $z=0$. Use of integral 3.323.2 in Ref. \cite{GRADSTEIN} yields the expression
\begin{align}\label{ROTATING}
\begin{split}
\psi_f(x,t',z) = &\sqrt{\frac{q_{\rm eff}}{z+q_{\rm eff}}}e^{-t^{\prime 2}/t_0^2}\exp\left[\frac{ik_0x^2}{2\left(z+q_{\rm eff}\right)}\right]\\
&\times\exp\left[\frac{\left(\frac{2t'px}{t_0^2}+\frac{2i(t'p)^2 z}{k_0 t_0^4}\right)q_{\rm eff}}{z+q_{\rm eff}}\right],
\end{split}
\end{align}
where $1/q_{\rm eff}=-1/f + 2i/k_0X_{0,\rm eff}^2$ is an initial, effective complex beam parameter, and $1/X_{0,\rm eff}^2 =1/X_0^2 + p^2/t_0^2$ is an initial, effective width. Figure \ref{Fig2} shows ST intensity patterns as the pulse focuses and beyond in a particular example with $p>0$. At the focal plane, the tilt angle is always 90 degrees, and at the far field, approaches zero. Note that the actual rotation in the $z$-$x$ plane is counterclockwise, since $t'=t-z/c$, corresponding to the positive transverse OAM $pX_0^2/4f$ with $p>0$.

Partially coherent rotating pulses have recently analyzed in \cite{HYDE}, where a sophisticated scheme for their generation using a Fourier transform pulse shaper with a spatial light modulator at the Fourier plane is proposed, and its transverse OAM noticed. It follows from our analysis that a simple prism tilting the pulse and focusing produces the rotating pulse, whose transverse OAM can be controlled by the focal length, tilt parameter and transversal size according to (\ref{TOAMR}).

Also, the above focused tilted Gaussian pulse has previously been used in experiments because of its important properties at the focal plane, particularly in high-harmonic generation experiments for the formation of the so-called attosecond lighthouses \cite{VICENTI_PRL,QUERE_JOP,AUGUSTE_PRA}. The property that creates these lighthouses is a {\it rotating wavefront in time} at the focal plane due to a transverse chirp at that plane. Instead, we stress here the rotating pulse front with propagation distance as a manifestation of its transverse OAM. In addition, (\ref{ROTATING}) is an analytical expression valid at any propagation distance for such interesting field that includes all above phenomena, namely, temporal wavefront rotation at the focal plane and pulse-front rotation on
propagation, excluding only the pulse-front curvature accrued during the act of focusing.

\subsection{Canonical STOV} 
\begin{figure}[t]
\includegraphics*[width=8.9cm]{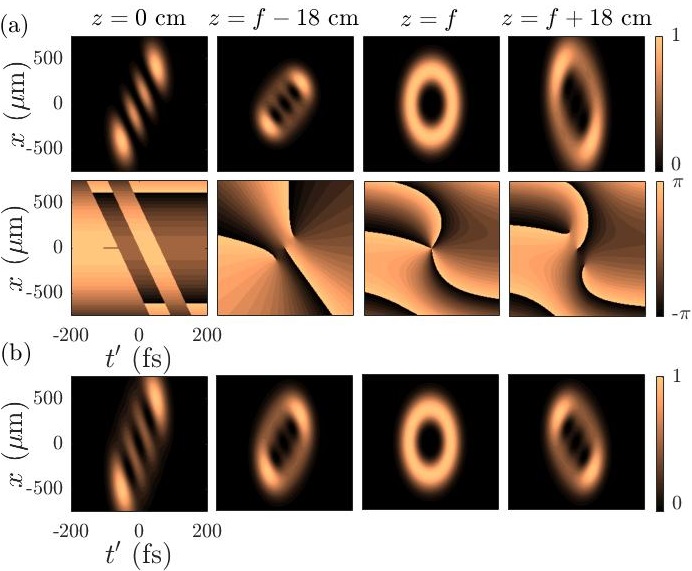}
\caption{\label{Fig3} (a) Elliptical STOV produced at the focal plane of a lens of focal length $f=50$ cm with the illumination in (\ref{BAND0}) of carrier frequency $\omega_0=2.5$ rad/fs, $t_0=50$ fs, $X_0=0.4$ mm, and $n=3$. Top: Intensity. Bottom: Phase of $\psi$. (b) Intensity of the STOV in Eq. (\ref{DEBYE}) with $x_0=2f/k_0X_0=0.3$ mm. In (a) and (b) the intensity is normalized to the peak intensity at each distance.}
\end{figure}

Let now the illumination be
\begin{equation}\label{BAND0}
\psi(x,t)= e^{-\frac{t^2}{t_0^2}} e^{-\frac{x^2}{X_0^2}}\frac{1}{2^n} H_n\left(\frac{t}{t_0}\pm \frac{x}{X_0}\right) ,
\end{equation}
where $H_n(\cdot)$ is the Hermite polynomial of order $n$. This illumination features $n+1$ tilted intensity lobes in the $t$-$x$ plane between the zeroes of the Hermite polynomial, as in the top right of Fig. \ref{Fig3}. The $1/2^n$ factor cancels the $2^n$ factor of the highest power term of the Hermite polynomial. This ``pre-conditioned" illumination with $n=1$ and $n=2$ has been shown \cite{CHEN,CHEN2} to produce STOVs of topological charges $n=1$ and $n=2$ by numerical calculation of the field at the focal plane. Also, similar pre-conditioned tilted lobes are the output from a $4f$ pulse shaper with a spiral phase plate placed at the Fourier plane in one of the experiments in \cite{HANCOCK_OPTICA}, this output also producing an elliptical STOV at the focal plane (far field) of a lens \cite{HANCOCK_OPTICA}. 

Below we provide an analytical description of the focused field at any distance from the focusing system producing elliptical STOVs of arbitrary charge $n$ at the focal plane, and evaluate its tranverse OAM. As discussed bellow, this focusing problem differs from the diffraction problem in \cite{HANCOCK_PRL} for topological charge $n=1$ and in \cite{PORRAS_OL} for arbitrary $n$, where the diffraction of a prescribed elliptical STOV from a waist plane is studied.

Being real, the illumination (\ref{BAND0}) does not carry transverse OAM. After focusing, (\ref{FTOAM2}) [or (\ref{FTOAM}) since the pulse front curvature contribution to the transverse OAM vanishes] yields the result
\begin{equation}\label{JSTOV}
\frac{J_{y,f}^{(i)}}{W}= \pm \frac{n X_0 t_0}{ 4f},
\end{equation}
which depends only on the focal length, duration and transversal size of (\ref{BAND0}), and will be better understood after examining the elliptical STOV at the focal plane. The change from zero to (\ref{JSTOV}) shows that it is just focusing that imparts the transverse OAM for the formation of the STOV, and therefore pre-conditioning systems other than the 4f pulse shaper can create the STOV, for example, a spatial light modulator to create the lobes and a prism to tilt them.

Using again (\ref{FRESNEL}) with $\psi(x,t)$ in (\ref{BAND0}), the resulting integral can be identified, after some changes of variables, with integral 7.374.8 in Ref. \cite{GRADSTEIN}, and the focused field written as
\begin{equation}\label{STOVZ}
\begin{split}
&\psi_f(x,t',z)=e^{-\frac{t^{\prime 2}}{t_0^2}} \left[\frac{q_0}{q(z)}\right]^{\frac{1}{2}} e^{\frac{ik_0x^2}{2q(z)}} \left[\frac{q_0}{q(z)}\left(1-\frac{z}{f}\right)\right]^{\frac{n}{2}}\\
&\times\frac{1}{2^n} H_n \left\{\left[\frac{q(z)}{q_0}\frac{1}{\left(1-\frac{z}{f}\right)}\right]^{\frac{1}{2}}\!\!\left(\frac{t'}{t_0}\pm \frac{x}{X_0}\frac{q_0}{q(z)}\right)\right\},
\end{split}
\end{equation}
where $1/q_0 = -1/f + 2i/k_0X_0^2$ and $q(z)=q_0+z$. Some ST intensity and phase profiles at different propagation distances are shown in Fig. \ref{Fig3}(a) in a particular example. The $n$ $\pi$-step tilted lines at the zeroes of the Hermite polynomial immediately form $n$ single-charged ST punctual phase singularities that merge in a single $n$-charged elliptical vortex at the focal plane, which further splits into $n$ unit-charged vortices up to the far field.

At the focal plane, $z=f$, the factor in the square bracket in the first row becomes zero and cancels all terms of the Hermite polynomial except the highest power term. Equation (\ref{STOVZ}) then reduces to
\begin{equation}\label{FOCUS}
\begin{split}
\psi_f(x,t',f ) & =  e^{-\frac{t^{\prime 2}}{t_0^2}} \left(\frac{-iX_0}{x_0}\right)^{\frac{1}{2}} e^{\frac{ik_0 x^2}{2f}} e^{-\frac{x^2}{x_0^2}}\\
&\times\left(\frac{t'}{t_0}\mp i\frac{x}{x_0}\right)^n,
\end{split}
\end{equation}
where $x_0= 2f/k_0X_0$ is the focal Gaussian width. This equation indeed represents an elliptical STOV of topological charge $n$ and ellipticity $\gamma=ct_0/x_0$. Following \cite{PORRAS_PIERS} or \cite{HANCOCK_PRL}, the intrinsic transverse OAM of a STOV can be evaluated from the topological charge $n$ at the plane where it is elliptical as
\begin{equation}\label{JSTOV2}
\frac{J_y^{(i)}}{W} = \pm \frac{\gamma}{2} \frac{n}{\omega_0} ,
\end{equation}
Using that $x_0=2f/k_0X_0$, Eq. (\ref{JSTOV2}) is immediately seen to coincide with Eq. (\ref{JSTOV}), that is, the transverse OAM of the elliptical STOV at the focal plane is that imparted by the lens.

\section{Discussion}

\subsection{Focusing to versus diffraction of  canonical STOVs}

In the second example above we have deliberately chosen a loose focusing geometry in Fig. \ref{Fig3}(a) with $Z_R/f=1.33$ to make it evident that our focusing problem is different from the diffraction problem studied in \cite{HANCOCK_PRL} and \cite{PORRAS_OL}, where collimated, elliptical STOVs are prescribed at a ``focus" and then they are propagated freely, i.e., diffracted forwards and backwards.

Figure \ref{Fig3}(b) shows the elliptical STOV of Ref. \cite{PORRAS_OL}, or given by Eq. (\ref{DEBYE}) below, of the same $x_0$, $t_0$ and $n$ at the same axial position as in Fig. \ref{Fig3}(a) for focusing, and their backwards and forwards propagated fields. In (a) for actual focusing, the focused field is not symmetric with respect to the focal plane, whilst in (b) the field diffracts symmetrically with respect to it. In (a) the waist [of the Gaussian beam enveloping the STOV in Eq. (\ref{STOVZ})] is located $18$ cm before the focal plane because of the relevant focal shift $f[(f^2/Z_R^2)/(1+(f^2/Z_R^2)]$ \cite{SELF,CARTER}, but in (b) the waist and the focal plane coincide. Indeed, the elliptical STOV in (a) contains the wave front curvature factor $e^{ik_0 x^2/2f}$ in Eq. (\ref{FOCUS}), whilst the elliptical STOV in (b) does not.

The focusing problem we have addressed can only be approximated by the diffraction problem when $Z_R/f\gg 1$, i.e., when focal shift is negligible \cite{SELF,CARTER,COLIN}. Under this condition, Eq. \ref{STOVZ} can readily seen to approach Eq. (7) in \cite{PORRAS_OL} with any $n$, and in particular to Eq. (7) in \cite{HANCOCK_PRL} for $n=1$. Specifically, Eq. (\ref{STOVZ}) reduces to 
\begin{equation}\label{DEBYE}
\begin{split}
\psi(x,t',z) & \simeq  e^{-t^{\prime 2}/t_0^2}\left(\frac{-f}{p(z)}\right)^{\frac{1}{2}} e^{\frac{ik_0x^2}{2p(z)}}\left(\frac{z}{p(z)}\right)^{\frac{n}{2}} \\ 
&\frac{1}{2^n} H_n\left[\left(\frac{p(z)}{z}\right)^{\frac{1}{2}}\left(\frac{t'}{t_0} \pm \frac{x}{x_0}\frac{z_R}{p(z)}\right)\right]\,,
\end{split}
\end{equation}
where $p(z)=(z-f)-iz_R$, $z_R= k_0^2x_0^2/2$ is the Rayleigh distance of the focused pulse, and again $x_0= 2f/k_0X_0$. Equation (\ref{DEBYE}) is the same as Eq. (7) in \cite{PORRAS_OL} except for a global amplitude to match the actual amplitude at the focal plane. Thus, as already pointed out in Ref. \cite{PORRAS_OL}, Eq. \ref{DEBYE}, and Eq. (7) in \cite{HANCOCK_PRL} in particular, are approximations to focusing to canonical STOVs valid only for $Z_R/f\gg 1$, while our Eq. (\ref{STOVZ}) is valid with any focusing geometry.

With regard to the intrinsic transverse OAM, we have demonstrated that it is imparted by the focusing system to the OAM-free illumination in Eq. (\ref{BAND0}). Instead, in the diffraction problems in \cite{HANCOCK_PRL} and \cite{PORRAS_OL}, the transverse OAM is theoretically introduced when an elliptical STOV is prescribed at the waist or focus, as \cite{HANCOCK_PRL} demonstrates that an elliptical STOV of charge $n$ carries the transverse OAM $(\pm \gamma/2)(n/\omega_0)$. There is no analysis of how it arrived there, and in particular of how a 4f pulse shaper (and the focusing lens afterwards) in \cite{HANCOCK_PRL} imparts the transverse OAM.

For completeness, we briefly consider the limit situation $f\rightarrow\infty$ in which the illumination in Eq. (\ref{BAND0}) is not focused but propagates freely undergoing diffraction. For $f\rightarrow\infty$ our Eq. (\ref{JSTOV}) yields zero transverse OAM. This is in agreement with the second example in \cite{PORRAS_OL}, Eqs.~(9-13) and Fig.~2, where the free-space diffraction of the same initial field without focusing is studied, and its intrinsic transverse OAM is calculated according to \cite{HANCOCK_PRL} to be zero.

\subsection{General fields with spatiotemporal intensity covariance}

\begin{figure}[t]
\includegraphics*[height=4.6cm]{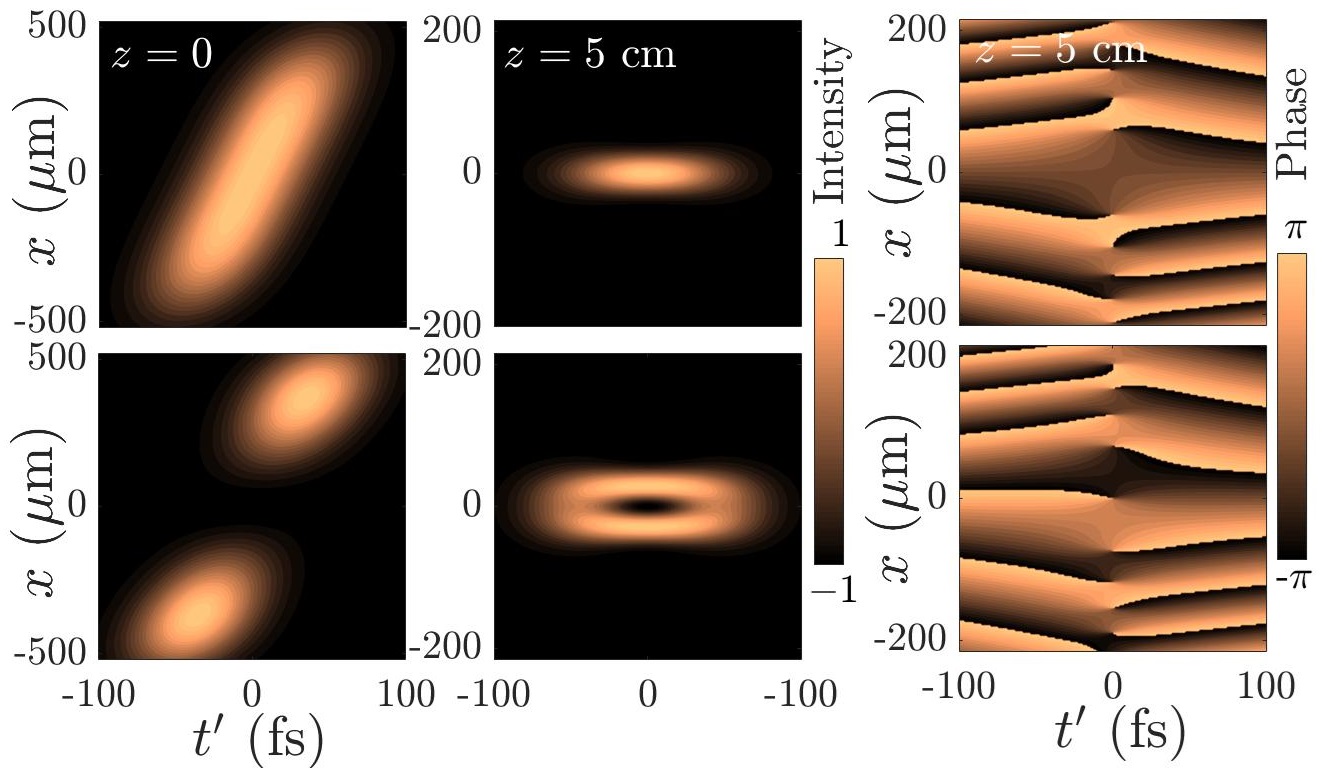}
\caption{\label{Fig4} Top: Rotating-type pulse with all parameters equal to those in Fig. \ref{Fig2}, except that the input field is $\psi=e^{-x^4/x_0^4}e^{-(t-px)^2/t_0^2}$ has a super-Gaussian transversal profile. Bottom: STOV-type pulse with all parameters equal to those in Fig. \ref{Fig2}, except that the input field is $\psi=(x/x_0)e^{-x^4/x_0^4}e^{-(t-px)^2/t_0^2}$ with a zero amplitude line at $x=0$. Left: intensity of illuminating field. Middle and right: intensity and phase at the focal plane.}
\end{figure}

The two examples in Sec. \ref{Examples} represent opposite situations where the imparted transverse OAM is manifested as a pure rotation of the pulse-front without ST singularities and as a formation of a canonical, elliptical STOV with an ST phase singularity. Other illuminations with ST intensity covariance produce mixed fields where the first phenomenon or the second dominate. For illuminating fields with no zeros in the amplitude other than the Gaussian tilted pulse, a rotating-type pulse is generally produced, which may be accompanied by many ST vortices of very low intensity far off-axis, as in Fig. \ref{Fig4} (top). A STOV-type pulse is produced when the amplitude has zero lines, as in Fig. \ref{Fig4}(bottom), where the central ST vortex is surrounded by an intense but imperfect ring, also accompanied by a myriad of low intensity ST vortices far off-axis, actually an infinite number of them at $t'=0$. The last example demonstrates that focusing of multi-lobe tilted structures with shapes other than Hermite-Gauss shape also create STOVs.

\subsection{Interaction with the focusing system}

It is also of interest to examine the (total) transverse OAM. Contrary to the intrinsic part, its value depends on the particular transverse axis. When the transverse $y$ axis $(x,z)=0$ passing through the center of the focusing system is chosen, the transverse OAM is given by \cite{PORRAS_PIERS}
\begin{equation}\label{TTOAM}
J_y = \frac{\varepsilon_0 z}{2k_0}\mbox{Im} \int \psi^\star\partial_x \psi  dx dt - \frac{\varepsilon_0}{2} \int |\psi|^2 x dx dt.
\end{equation}
It is obviously zero for an input collimated wave packet with $x_c=0$ (irrespective that $z=0$ or not). A similar analysis under the same conditions that led to (\ref{FTOAM}), leads now to the conclusion that the total transverse OAM continues to vanish after focusing. By conservation of the total angular momentum of the pulse-lens system, the pulse
does not transmit angular momentum to the lens. Indeed, in Fig. \ref{Fig1} the recoil momenta opposite to $p_x$ and $-p_x$ do not provide angular momentum to the lens, and the small amounts of axial momenta transferred from $p_z$ to the lens in the upper and lower parts provide opposite angular momenta and then zero net angular momentum. Focusing is then a zero-exchange interaction with regard angular momentum in which the null OAM of the pulse is split into an intrinsic part and an opposite extrinsic part.

\section{Conclusions}

To conclude, ST couplings prove to be extremely useful for generating structured light carrying transverse OAM. This fact has also been demonstrated recently in Ref. \cite{PORRAS_ARXIV}, where arbitrarily oriented vortices are produced from longitudinal vortices with spatial chirp. In this work
we have proposed a simple procedure to impart transverse OAM by focusing an ST coupled field the ST intensity pattern of which presents covariance in space and time.

Except in \cite{HYDE}, research on transverse OAM has focused to the ST phase singularities of STOVs. We have evaluated analytically the field of coherent, rotating pulses in full space and time, which can be of interest for the analysis of their interaction with matter, e.g., in high harmonic and attosecond pulse generation \cite{VICENTI_PRL,QUERE_JOP,AUGUSTE_PRA}, and evaluated its transverse OAM as a function of the input transversal size, tilt parameter and focal length that generate the rotating pulse.

STOVs can also be created by focusing specific profiles, where we now show that the covariance of the intensity in space and time is the key property which ensures the transverse OAM upon focusing. Hermite-Gauss, ST tilted multi-lobe profiles focus to a perfect STOV, whose focusing field is analytically described, and the transverse OAM is shown to be that imparted by the focusing system. Other multi-lobe profiles also focus to more or less perfect STOVs with the OAM imparted by the focusing system. To clarify these nuances, we also showed the focusing properties of cases in-between the simple tilted pulse and the STOV that always contain transverse OAM but show varying levels of vortex structure. We finally note that the magnitude of transverse OAM in the limit cases of the rotating pulse and STOV are, under similar focusing conditions and input duration and size, of the same order of magnitude, implying that pulses with transverse OAM but without ST phase singularities may be just as useful for future applications.

\section*{Funding}
Ministerio de Ciencia e Innovación (PID2021-122711NB-C21); Horizon 2020 Framework Programme (801505).

\section*{Acknowledgments}
This work has been partially supported by the Spanish Ministry of Science and Innovation, Gobierno de España, under Contract No. PID2021-122711NB-C21. S.W.J. has received funding from the European Union’s Horizon 2020 research and innovation program under the Marie Skłodowska-Curie grant agreement No 801505.





\begin{thebibliography}{99}
\bibitem{WAN} C. Wan, A. Chong, and Q. Zhan,  ``Optical spatiotemporal vortices," eLight {\bf 3}, 11 (2023).
\bibitem{HANCOCK_OPTICA} S.W. Hancock, S. Zahedpour, A. Goffin, and H.M. Milchberg, ``Free-space propagation of spatiotemporal optical vortices," Optica {\bf 6}, 1547--1553 (2019).
\bibitem{CHONG} A. Chong, C. Wan, J. Chen, and Q. Zhan, ``Generation of spatiotemporal optical vortices
with controllable transverse orbital angular momentum," Nat. Photonics {\bf 14}, 350 (2020).
\bibitem{CAO} Q. Cao, J. Chen, K. Lu, Ch. Wan, A. Chong, and Q. Zhan ``Non-spreading Bessel spatiotemporal optical vortices," Science Bulletin {\bf 67}, 133 (2022).
\bibitem{HUANG} J. Huang, J. Zhang, T. Zhu, and Z. Ruan, ``Spatiotemporal differentiators generating optical vortices with transverse orbital angular momentum and detecting sharp change of pulse envelope," Laser Photonics Rev. {\bf 16}, 2100357 (2022).
\bibitem{TORQUING} S.W. Hancock, S. Zahedpour, A. Goffin, H.M. Milchberg, ``Spatiotemporal torquing of light," arXiv:2307.01019 (2023).
\bibitem{VICENTI_PRL} H. Vincenti and F. Qu\'er\'e, ``Attosecond Lighthouses: How To Use Spatiotemporally Coupled Light Fields To Generate
Isolated Attosecond Pulses," Phys. Rev. Lett. {\bf 108}, 113904 (2012).
\bibitem{QUERE_JOP} F. Qu\'er\'e, H. Vincenti, A. Borot, S. Monchoc\'e1, T.J. Hammond,
K.T. Kim, J.A. Wheeler, Ch. Zhang, T. Ruchon, T. Auguste, J.F. Hergott, D.M. Villeneuve, P.B. Corkum,
and R. Lopez-Martens, `` Applications of ultrafast wavefront rotation
in highly nonlinear optics," J. Phys. B: At. Mol. Opt. Phys. {\bf 47}, 12004 (2014).
\bibitem{AUGUSTE_PRA} T. Auguste, O. Gobert, T. Ruchon, and F. Qu\'er\'e, ``Attosecond lighthouses in gases: A theoretical and numerical study," Phys. Rev. A {\bf 93}, 033825 (2016).
\bibitem{HYDE} M.W. Hyde, ``Twisted space-frequency and space-time partially coherent beams," Sci. Rep. {\bf 10}, 12443 (2020).
\bibitem{FANG} Y. Fang, S. Lu, Y. Liu, ``Controlling Photon Transverse Orbital Angular Momentum in High Harmonic Generation," Phys. Rev. Lett. {\bf 127}, 273901 (2021).
\bibitem{CHEN} J. Chen, C. Wan, A. Chong, and Q. Zhan, ``Subwavelength focusing of a spatio-temporal wave packet with transverse orbital angular momentum." Opt. Exp. {\bf 28}, 18472 (2020).
\bibitem{CHEN2}J. Chen, S. Kuai, G. Chen, L. Yu, and Q. Zahn, ``Dynamical Modulation of Transverse Orbital Angular Momentum in Highly Confined Spatiotemporal Optical Vortex," Photonics {\bf 10}, 148(2023).
\bibitem{PORRAS_PIERS} M.A. Porras, ``Transverse orbital angular momentum of spatiotemporal optical vortices," 	Prog. Electromagn. Res. {\bf 177}, 95--105 (2023).
\bibitem{HANCOCK_PRL} S.W. Hancock, S. Zahedpour, and H.M. Milchberg, ``Mode Structure and Orbital Angular Momentum of Spatiotemporal Optical Vortex Pulses," Phys. Rev. Lett. {\bf 127}, 193901 (2021).
\bibitem{BLIOKH_PRA} K.Y. Bliokh, ``Orbital angular momentum of optical, acoustic, and quantum-mechanical
spatiotemporal vortex pulses," Phys. Rev. A, {\bf 107}, L031501 (2023).
\bibitem{SELF} S.A. Self, ``Focusing of Spherical Gaussian Beams," Appl. Opt. {\bf 22}, 658--661 (1983).
\bibitem{CARTER} W.H. Carter, ``Focal shift and concept of effective Fresnel number for a Gaussian laser beam," Appl. Opt. {\bf 21}, 1989--1994 (1983).
\bibitem{COLIN} C.J.R. Sheppard, ``Validity of the Debye approximation," Opt. Lett. {\bf 25}, 1660--1662 (2000).
\bibitem{BOR93} Z. Bor, B. R{\'a}cz, G. Szab{\'o}, M. Hilbert, and H. A. Hazim, ``Femtosecond pulse front tilt caused by angular dispersion," Optical Engineering {\bf 32}, 2501--2504 (1993).
\bibitem{PRETZLER00} G. Pretzler, A. Kasper, and K. J. White, ``Angular chirp and tilted light pulses in {CPA} lasers," Applied Physics B {\bf 70}, 1--9 (2000).
\bibitem{GRADSTEIN} I.S. Gradshteyn and I.M. Ryzhik, {\it Table of Integrals, Series, and Products,} Fifth Edition, Academic Press (1994).
\bibitem{PORRAS_OL} M.A. Porras, ``Propagation of higher-order spatiotemporal vortices," Opt. Lett. {\bf 48}, 367--370 (2023).
\bibitem{PORRAS_ARXIV} M.A. Porras and S.W. Jolly, ``Control of vortex orientation of ultrashort optical pulses
using spatial chirp," 
Opt. Lett. {\bf 48}, 6448-6451 (2023).
\end{thebibliography}
\end{document}